# Influence of Ni doping on critical parameters of PdTe superconductor

*Reena Goyal[1,2], Rajveer Jha[1], Brajesh Tiwari[3], Ambesh Dixit[4] and V.P.S. Awana[1,*]*

[1]*CSIR-National Physical Laboratory, Dr. K. S. Krishnan Marg, New Delhi-110012, India*
[2]*Academy of Scientific and Innovation Research, NPL, New Delhi-110012*
[3]*Sardar Vallabhbhai National Institute of Technology, Surat-395007, Gujarat, India*
[4]*Department of Physics, Indian Institute of Technology, Jodhpur, Rajsthan, India*

**Abstract**

We report the effect of Ni doping on superconductivity of PdTe. The superconducting parameters like critical temperature ($T_c$), upper critical field ($H_{c2}$) and normalized specific-heat jump ($\Delta C/\gamma T_c$) are reported for Ni doped $Pd_{1-x}Ni_xTe$. The samples of series $Pd_{1-x}Ni_xTe$ with nominal compositions x=0, 0.01, 0.05, 0.07, 0.1, 0.15, 0.2, 0.3 and 1.0 are synthesized via vacuum shield solid state reaction route. All the studied samples of $Pd_{1-x}Ni_xTe$ series are crystallized in hexagonal crystal structure as refined by Rietveld method to space group $P6_3/mmc$. Both the electrical resistivity and magnetic measurements revealed that $T_c$ decreases with increase of Ni concentration in $Pd_{1-x}Ni_xTe$. The magneto-transport measurements suggest that flux is better pinned for 20% Ni doped PdTe as compared to other compositions of $Pd_{1-x}Ni_xTe$. The effect and contribution of Ni 3d electron to electronic structure and density of states near Fermi level in $Pd_{1-x}Ni_xTe$ are also studied using first-principle calculations within spin polarized local density approximation. The overlap of bands at Fermi level for NiTe is larger as compared to PdTe. Also the density of states just below Fermi level (in conduction band) drops much lower for PdTe than as for NiTe. Summarily, Ni doping in $Pd_{1-x}Ni_xTe$ superconductor suppresses superconductivity moderately and also Ni is of non magnetic character in these compounds.

Key words: Superconductivity, Magneto transport, Heat capacity, PdTe superconductor, Ni doping

PACS number(s): 74.70.-b, 74.25.F-, 74.10.+v, 74.25.Ha

[*]**Corresponding Author**
Dr. V. P. S. Awana: E-mail: awana@mail.nplindia.org
Ph. +91-11-45609357, Fax-+91-11-45609310
Homepage awanavps.webs.com

**Introduction**

The discovery of new superconductors always attracted enormous interest from both experimental and theoretical condensed matter physics community. For the meantime, the already discovered superconductors keep on motivating to understand the underlying physics of them. The role of magnetic impurities in known superconductors has been of great interest for a long time [1-4]. For example the magnetic Mn impurities in $Ba_{0.5}K_{0.5}Fe_2As_2$ and $Ba(Fe_{1-y}Co_y)_2As_2$ systems showed strong suppression of $T_c$ [5-7], while $T_c$ is nearly unchanged in Mn-doped $FeSe_{0.5}Te_{0.5}$ superconductors [8]. The non magnetic Zn doped $BaFe_{1.89-2x}Zn_{2x}Co_{0.11}As_2$ compounds showed that the $T_c$ decreases rapidly with increasing Zn doping level [9], but the superconducting state is quite robust for $Fe_{1-y}Zn_ySe_{0.3}Te_{0.7}$ compound [10]. In case of high $T_c$ Cuprates significant decrease in $T_c$ was observed with Cu site Zn doping [11-13]. Studying the effect of both magnetic and non-magnetic impurities on known superconductors had been of prime interest for over the years [1-13].

Recently, PdTe superconductor got attention [14, 15]. These works prompted us to study the effect of magnetic Nickel on bulk polycrystalline PdTe superconductor. The effects of magnetic impurities and the possibility of magnetic ordering in BCS type conventional PdTe superconductor could provide better understanding of superconductivity. Generally, it has been believed that the conduction electrons cannot be ordered both magnetically and superconducting due to strong spin scattering [16, 17]. On the other hand, cooper pairs are formed in Cuprates and Iron based compounds possibly through spin fluctuations and superconductivity occurs after suppressing the magnetic ordering by chemical doping or the application of hydrostatic pressure [18-20]. The electron-phonon coupling as proposed in BCS theory failed to explain the superconductivity in Cuprate and Iron based materials [21]. The superconductivity in high $T_c$ Cuprates is induced from electronic charge carriers doping in antiferromagnetic Mott insulating phase [21–23]. There is a hypothetical possibility of the magnetic excitations being replacing phonons in exotic high $T_c$ superconductors [23]. On the other hand, there are some examples for the coexistence of superconductivity with either ferromagnetic or anti-ferromagnetic ordering in $UGe_2$, URhGe, UCoGe, $MgCNi_3$ and $RuSr_2GdCu_2O_8$ etc [24-28]. As far as the coexistence of superconductivity and magnetism is concerned, there is no concrete explanation to understand the interaction between superconducting and magnetic order parameters. In some experimental reports, it has been suggested that the $T_c$ decreases linearly with increasing magnetic impurity concentration in superconducting systems [6-9]. The decrease in $T_c$ of bulk lanthanum by rare-earth impurities depends on the spin of the impurity atoms rather than on their magnetic moment, which has been reported by Matthias in a detailed study [29-33].

Keeping in view the importance of the impact of magnetic ions doping in various superconductors, we report here on synthesis and characterisation of $Pd_{1-x}Ni_xTe$ (0≤ x ≤1) series. For pristine PdTe, our results are a short of approval for the only other report [14] available in literature on superconductivity at 4.5K in PdTe, besides our previous work [15]. The difference, which we feel is important that the earlier report is on tiny (15μm) single crystals [14], the present one is on the other hand on polycrystalline bulk samples. Also the magneto-heat capacity was added [15], which is useful in not only probing the bulk superconductivity but the order parameter as well.

The X-ray diffraction (XRD) measurements revealed that the Pd gets substituted by Ni in the parent hexagonal phase (space group $P6_3/mmc$) of PdTe. The superconducting transition temperature $T_c$ of $Pd_{1-x}Ni_xTe$ is studied by resistivity measurements using QD-PPMS down to 2 K under different magnetic fields. The Heat capacity measurements for $Pd_{0.99}Ni_{0.01}Te$ are also presented and analysed. Ni doping in $Pd_{1-x}Ni_xTe$ decreases superconductivity moderately, and the reason behind is that Ni is of non magnetic nature in PdTe. The Ni(3d) and Te(sp) orbital possible strong hybridisation might be the reason behind non magnetic nature of Ni in $Pd_{1-x}Ni_xTe$. Detailed first principal density functional calculations revealed that Ni affects the Te-p orbitals, resulting in suppression of superconducting transition temperature. Interestingly, to best of our knowledge this is the first study on Ni substitution at Pd site in PdTe superconductor.

**Experimental and Computational Details**

The bulk polycrystalline samples of series $Pd_{1-x}Ni_xTe$ (0≤ x ≤1) were synthesized by the solid state route via vacuum encapsulations. The required elements Pd (3N), Te (4N) and Ni (4N) were grinded in a stoichiometric ratio in the argon filled glove box. The powders were pelletized by applying uniaxial stress of 100kg/cm$^2$ and vacuum sealed (<10$^{-3}$Torr) in quartz tubes. The sealed encapsulated quartz tubes were kept in a box furnace and then heated to 750°C (rate 2°C/min) for 24 hours and cooled down to room temperature naturally. The obtained samples were dense and shiny black. For the structural analysis, X-ray diffraction (XRD) is done at room temperature using CuK$_α$ radiation of wavelength 1.5418Å. Magneto resistance measurements were performed by four probe technique in an applied field in quantum design Physical Property measurements system (PPMS-14 Tesla) - down to 2K. Specific heat measurements were also carried at same facility. Magnetic measurements were performed in MPMS system. DC magnetization with temperature variation was carried in both zero field and field cooled modes.

We also performed density functional calculations to see the role of Ni doping on electronic structure of PdTe. These calculations were carried out within generalized gradient approximation (GGA) of Perdew, Burke and Ernzerhof, as implanted in the Vienna Ab- initio Simulation Package (VASP), to compute the ground state electronic band structure and density of states [34].

**Results and Discussion**

In Fig.1, the observed and Rietveld fitted room temperature XRD patterns of the $Pd_{1-x}Ni_xTe$ samples are shown. All the samples are well fitted with the space group *P6₃/mmc*, suggesting complete solubility of Ni in PdTe. It can be observed from Fig.1 that (100) crystallographic plane around $2\theta=24.82^o$ is being suppressed with increase in concentration of Ni. On the other hand the (002) plane at $2\theta=33.07^o$ appears only above x=0.1. The lattice parameters for PdTe are; $a=b=4.153(2)$Å and $c=5.673(5)$Å and for NiTe $a=b=3.941(2)$Å and $c=5.3632(5)$ Å as obtained by fitting to *P6₃/mmc* space group. As seen from upper panel of Fig. 2, the Rietveld refined lattice parameters $a=b$ and $c$ and unit cell volume ($V$) are consistently decreasing with increasing Ni-doping fraction in $Pd_{1-x}Ni_xTe$ series. Almost linear shrinkage of the unit cell volume of $Pd_{1-x}Ni_xTe$ with x indicates complete substitution of Pd by Ni in PdTe, suggesting the increase of chemical pressure. The lower panel of Fig. 2 shows that the lattice parameters $a$ and $c$ decrease simultaneously. Although, we could not carry out the elemental analysis, the linear decrease in lattice parameters indicates successful substitution of smaller ion Ni at Pd site in $Pd_{1-x}Ni_xTe$. For brevity, we took the nominal x value as Ni content in studied $Pd_{1-x}Ni_xTe$. The chemical pressure may play an important role on the superconductivity of parent PdTe compound. For Pd doped FeTe compound, it has been reported that the negative chemical pressure as well as doping induced structural phase transition occurs from tetragonal to hexagonal phase [35].

Fig. 3 represents the ac susceptibility for the all superconducting samples of series $Pd_{1-x}Ni_xTe$ (0≤ x ≤0.2). Both the real (*M'*) and imaginary (*M''*) parts of the ac magnetic susceptibility measurements are carried out at an amplitude of 10Oe and frequency 333Hz down to 2K. The M' showed a sharp transition to diamagnetism ($T_c$) at around 4.5K, confirming the bulk superconductivity in pristine PdTe sample. In contrast, M" exhibited a sharp single positive peak around the same temperature, indicating strong superconducting grains coupling in PdTe superconductor. With increasing Ni concentration in $Pd_{1-x}Ni_xTe$, the $T_c$ shifts monotonically to lower temperatures from 4.5K (x=0.0) to 2.5K for x= 0.20 sample. The Ni substitutions above 20% in PdTe did not show superconductivity in the studied temperature range down to 2K.

Fig. 4 shows the magnetization isotherms at 2K to estimate the lower critical field ($H_{c1}$) values of superconducting $Pd_{1-x}Ni_xTe$ samples. With increasing magnetic field from zero, the magnetization increases linearly up to $H_{c1}$ signifying the diamagnetic character. For PdTe the value of magnetization starts increasing above the magnetic field $H_{c1}$, reaches to zero i.e., at upper critical field $H_{c2}$ and becomes positive above an applied field of 1kOe. The estimated $H_{c1}$ values are 200Oe, 160Oe, 51Oe, 41Oe, and 32Oe for $Pd_{1-x}Ni_xTe$; x=0, 0.01, 0.05, 0.07 and 0.1, respectively. Clearly, the $H_{c1}$ of $Pd_{1-x}Ni_xTe$ series decreases with increase in Ni content.

Temperature dependence of electrical resistivity for $Pd_{1-x}Ni_xTe$ (0≤ x ≤1.0) series in the temperature range 300-2K as shown in Fig. 5a. It is important to check the variation of normal state resistivity with Ni doping, as the same hosts the superconductivity at low temperatures. Apparently, normal state electrical resistivity of all the samples increases with temperature, albeit with metallic behaviour. Fig. 5b shows ρ-$T$ plots of Ni doped PdTe superconducting compounds in the temperature range 6-2K. The onset resistivity increases while superconducting transition temperature ($T_c$) decreases with increasing Ni substitution at Pd site. This trend is shown in Fig. 6 in terms of $T_c^{onset}$ vs Ni content plot for $Pd_{1-x}Ni_xTe$ superconducting samples. Here one can see that the $T_c^{onset}$ is nearly unchanged for x=0.01 sample and the same decreases rapidly for higher Ni content samples. None of the samples showed superconducting transition down to 2K for x ≥ 0.3. Fig. 7 shows the variation of residual resistivity $\rho_o$ and the residual resistivity ratio RRR (ratio of resistivity at 300 K to the extrapolated resistivity at zero K) with Ni concentration. The residual resistivity $\rho_o$ has been calculated by the fitting of electrical resistivity using equation $\rho=\rho_o+ AT^2$, where A is the slope of the graph, shown in inset of Fig.7. RRR is found to decrease with Ni content, suggesting increased scattering of electrons with Ni doping. In our case the residual resistivity increases monotonically up to the doping level of 20%, and later rapidly increases as the superconductivity disappears. The suppression of $T_c$ in $Pd_{1-x}Ni_xTe$ system may result from the change of the charge carrier density along with the impurity scattering.

Fig. 8 (a-g) demonstrates the temperature dependence of electrical resistivity under various magnetic fields in the temperature range 2-6K for superconducting $Pd_{1-x}Ni_xTe$ (x=0 to 0.2) samples. The $T_c^{onset}$ and $T_c^{(\rho=0)}$ decrease with increasing magnetic field and this superconducting behaviour has been observed for all the samples. Fig. 8h shows the upper critical field $H_{c2}$ corresponding to the temperatures where the resistivity drops to 90% of the normal state resistivity. The $H_{c2}(0)$ is estimated by using the conventional one-band Werthamer–Helfand–Hohenberg (WHH) equation, i.e., $H_{c2}(0)=-0.693T_c(dH_{c2}/dT)_{T=Tc}$. The solid lines are the ones being extrapolated to $T = 0K$, for 90% $\rho_n$ criteria of $\rho(T)H$ curve for $Pd_{1-x}Ni_xTe$ samples. The estimated $H_{c2}(0)$ values are 2.6kOe, 2.3kOe, 2.4kOe, 2.6kOe, 2.61kOe, 2.66kOe and 3kOe for $Pd_{1-x}Ni_xTe$ (x=0, .01, 0.05, 0.07, 0.1, 0.15 and 0.2) samples. The $H_{c2}(0)$ value for 20% Ni doped PdTe is significantly higher than the pristine sample, while its $T_c$(3K) is lower than the $T_c$(4.5K) of pristine PdTe sample. The upper critical field values $H_{c2}(0)$ estimated for all the samples are well within Pauli Paramagnetic limit which is defined as $\mu_oH_p=1.84T_c$[36].

To see the effect of magnetic doping on the electronic heat capacity of one of the superconducting compound $Pd_{0.99}Ni_{0.01}Te$, the low temperature specific heat has been recorded under different magnetic fields as shown in Fig 9(a). As mentioned in the experimental section the specific heat ($C_p$) measurements are carried out on Quantum Design (QD) PPMS with an instrument accuracy of 10nJ/K at 2K. Critical examination of heat capacity measurements made on a Quantum Design (QD) Physical Property Measurement System (PPMS) reported an accuracy of the same within 1 to 5%. The detailed inter comparison and data analysis is given in ref. 37.

. In absence of any applied field, the anomaly in specific heat ($C_p$) is observed at temperature 4.5K, which decreases to low temperature with magnetic field. The superconductivity anomaly is suppressed and not seen down to 2K at applied magnetic field of 1.5kOe. The specific heat is fitted to the expression $C_p(T) = \gamma T + \beta T^3 + \delta T^5$, where $\gamma$ is Sommerfeld coefficient, $\beta$ and $\delta$ are the phononic heat coefficients. Inset of Fig. 9 (a) represents $C_p/T$ as a function of $T^2$ which is used to find the electronic and phononic contribution to the specific heat. The obtained coefficients are $\gamma = 7.42$mJ mol$^{-1}$K$^{-2}$, $\beta = 0.8$mJ mol$^{-1}$K$^{-4}$ and $\delta = 0.0019$mJmol$^{-1}$K$^{-6}$. The Debye temperature ($\theta_D$) is 229.9K, which is calculated by using the relation $\theta_D = (234zR/\beta)^{1/3}$, where R is the Rydberg constant (8.314Jmol$^{-1}$ K$^{-1}$) and z is the number of atoms in the Ni doped PdTe unit cell. The Kadowaki–Woods ratio $A/\gamma^2$ is $8.7\times10^{-5}\mu\Omega$cm mol$^2$K$^2$J$^{-2}$, where A is evaluated by fitting of temperature dependent resistivity in previous section. Interestingly, the value of Kadowaki–Woods ratio for Pd$_{0.99}$Ni$_{0.01}$Te sample is in good agreement with transition metal systems [38]. The value of δ is so small that the data could be fitted well even without the $\delta T^5$ term. The fitting of $C_p(T)$ without $\delta T^5$ term gave γ = 6.666mJ mol$^{-1}$K$^{-2}$ and β = 0.884mJ mol$^{-1}$K$^{-4}$. This resulted in θ$_D$ as 223K and Kadowaki–Woods ratio $A/\gamma^2$ of $10.5\times10^{-5}\mu\Omega$cm mol$^2$K$^2$J$^{-2}$. In any case, both with and without $\delta T^5$ term, the fitting resulted in close θ$_D$ and Kadowaki–Woods ratio values. As far as the goodness of fitting is concerned the same is 0.9998(6) for $\delta T^5$ term and 0.9997(6) without the same. For different magnetic fields, the variation of electronic specific heat $C_e/T$ ($C_e=C_p-\beta T^3- \delta T^5$) with temperature is represented in Fig.9b. It is observed that specific heat jump decreases with magnetic field along with superconducting transition temperature. The electronic specific heat is calculated by subtracting phononic part from total specific heat. This plot is used to determine jump in electronic heat capacity at superconducting $T_c$. It can be seen that the magnitude of the jump (ΔC) at $T = T_c$ is 10.59mJ/molK$^2$, and the value of the normalized specific-heat jump, (ΔC/γ$T_c$) is 1.42, nearly equal to the BCS weak-coupling limit, i.e., 1.43. This value is slightly larger than as for PdTe, being reported (1.33) in our previous work [15]. Interestingly, the jump value even for pristine PdTe is 1.67 in ref. 14 and 1.33 in ref. 15. As mentioned by us in an earlier report [15], the jump value depends upon the superconducting volume fraction and disorder may also affect the same, hence it is difficult to comment and compare the exact values. The specific-heat jump, (ΔC/γ$T_c$) for Pd$_{0.99}$Ni$_{0.01}$Te along with our earlier report [15] on pristine PdTe, suggests that the superconductivity of these compounds is within the BCS coupling limit.

Fig. 10 (a-e) shows the near Fermi energy electric band structure of Pd$_{1-x}$Ni$_x$Te; x=0.0, 0.25, 0.5, 0.75 and 1.0, calculated using first principle within density functional approximations, as implemented in VASP. A 2x2x2 PdTe superstructure was used for the calculation of Ni substituted Pd$_{1-x}$Ni$_x$Te samples. The structures were relaxed until force on each atom was less than 0.01eV/Å for all these compositions. The starting structures were taken from experimental lattice parameters, being estimated using Rietveld refinement. Apart from the angles α=89.3° and β=91.4° for doped samples, other lattice parameters are nearly close to experimental values. The details of the calculated parameters are listed in Table 2. Though small but this changes the crystal symmetry, which leads to large number of energy bands in Ni-substituted samples. These changes in lattice parameters are not discernible from x-ray diffraction measurements. Except for the overlap of bands at Fermi level there are no other changes for PdTe and NiTe compounds. The overlap for NiTe is larger as compared to PdTe across the Fermi level. Also worth noticing that density of states just below Fermi level (in conduction band) drops much lower for PdTe as compared to NiTe, consistent with electronic band overlap, as can be seen in Fig 10. This instability at Fermi level may be responsible for superconducting gap in PdTe and lowering of superconducting transition temperature $T_c$ with increasing Ni-doping fraction in PdTe. The residual resistivity is much lower for PdTe than that of NiTe, as can be seen in Fig. 7. Current carriers of these studied compounds have Fermi liquid like behaviour at low temperatures. This suggests that the nature of bands near Fermi level is important to explain the residual resistivity in conjunction with impurity scattering. Spin polarized density of states (DOS) per unit cell are shown in the right panel of Fig. 10 and non-zero density of states at Fermi level (Fermi level is set to zero) has been observed for all $Pd_{1-x}Ni_xTe$ solid solutions. In fact, to probe the role of Ni-doping in superconducting PdTe and non-superconducting NiTe compounds further studies on the nature of Fermi surface and electronic bands are yet warranted. Ours are though preliminary observations, but are certainly thought provoking and deserves further investigation.

**Conclusion**

In summary, we successfully synthesized Ni doped PdTe compounds, the XRD pattern for the $Pd_{1-x}Ni_xTe$ (0≤ x ≤1.0). Superconductivity ($T_c$) decreases with increase in Ni content and is completely disappeared at above 20% Ni doping. Interestingly, Ni is found to be of non magnetic nature in $Pd_{1-x}Ni_xTe$, and hence the $T_c$ depression is mainly due to disorder alone. The $H_{c2}(0)$ value for 20% Ni doped PdTe is significantly higher than the pristine PdTe sample, suggesting possible pinning. The value of the normalized specific-heat jump ($\Delta C/\gamma T_c$) of 1.42 is estimated from the analysis of specific heat data of $Pd_{1.99}Ni_{0.01}Te$, suggesting a simple BCS weak-coupling limit. Worth mentioning is the fact that, this is first study on Ni doping in PdTe superconductor, which may yet have loose ends and further investigations may though be desired.

**Acknowledgement**

The authors would like to thank the Director of NPL-CSIR India for his encouragement. This work is financially supported by a DAE-SRC outstanding investigator award scheme on the search for new superconductors. RG thanks UGC, India, for providing her research fellowship.

Table 1: Normal state resistivity fitted parameters evaluated from equation $\rho=\rho_0+AT^2$

| $Pd_{1-x}Ni_xTe$ | $\rho_0(\mu\,\Omega\text{-cm})$ | $A(\mu\,\Omega\,\text{cm}-K^{-2})$ |
|---|---|---|
| x=0 | 5.34 | $6.10038\times10^{-9}$ |
| x=0.01 | 7.68 | $4.82482\times10^{-9}$ |
| x=0.05 | 15.69 | $4.96768\times10^{-9}$ |
| x=0.07 | 17.02 | $3.93908\times10^{-9}$ |
| x=0.1 | 25.81 | $4.09503\times10^{-9}$ |
| x=0.15 | 28.57 | $3.03277\times10^{-9}$ |
| x=0.2 | 43.51 | $3.74791\times10^{-9}$ |
| x=0.3 | 83.34 | $5.26336\times10^{-9}$ |
| x=1 | 242.42 | $6.93269\times10^{-10}$ |

Table 2: Lattice parameters calculated from Density functional theory.

| $Pd_{1-x}Ni_xTe$ | $a$ (Å) | $b$ (Å) | $c$ (Å) |
|---|---|---|---|
| x=0 | 4.212(2) | 4.212(2) | 5.739(3) |
| x=0.25 | 4.216(2) | 4.191(1) | 5.574(2) |
| x=0.50 | 4.174(3) | 4.174(3) | 5.435(3) |
| x=0.75 | 4.180(2) | 3.987(2) | 5.449(1) |
| x=1 | 4.092(1) | 4.092(1) | 5.186(2) |

**Figure captions**

**Figures 1:** Experimental (red open circles) and Rieitveld refined (black solid line) room temperature x-ray diffraction patterns of $Pd_{1-x}Ni_xTe$ ($0 \leq x \leq 1$) compounds. The bottom (blue) lines correspond to the difference between the experimental and calculated data.

**Figures 2:** Nominal x dependence Rietveld fitted cell parameters $a(Å)$, $c(Å)$ and $V(Å^3)$ for $Pd_{1-x}Ni_xTe$ samples.

**Figures 3:** Temperature dependence of ac susceptibility for superconducting $Pd_{1-x}Ni_xTe$ (x=0, .01, 0.05, 0.07, 0.1, 0.15, 0.2 and 0.3) compounds.

**Figures 4:** Isothermal magnetization vs dc magnetic field in superconducting state at 2K for $Pd_{1-x}Ni_xTe$ (x=0, .01, 0.05, 0.07 and 0.1) compounds.

**Figures 5:** Temperature dependence of electrical resistivity of $Pd_{1-x}Ni_xTe$ (x=0, .01, 0.05, 0.07, 0.1, 0.15, 0.2, 0.3 and 1) series (a) in the temperature range 300-2K and (b) Zoomed part of the same in the superconducting region 6-2K.

**Figures 6:** Nominal x dependence of $T_c^{onset}$ (K) of superconducting $Pd_{1-x}Ni_xTe$ samples.

**Figures 7:** Nominal x dependence of residual Resistivity ($\rho_0$) and residual resistivity ratio ($\rho_{300}/\rho_0$) of $Pd_{1-x}Ni_xTe$ samples. Inset shows the method of $\rho(T)$ curve fitting in the relation $\rho=\rho_o+ AT^2$.

**Figures 8:** (a-g) Temperature dependence of electrical resistivity under various magnetic fields of superconducting $Pd_{1-x}Ni_xTe$ (x=0, .01, 0.05, 0.07, 0.1, 0.15 and 0.2) compounds. (h) Upper critical field ($H_{c2}$) as a function of temperature solid lines is linearly extrapolation of experimental data.

**Figures 9:** (a) Temperature dependence of heat capacity ($C_p$) under various magnetic fields of superconducting $Pd_{0.99}Ni_{0.01}Te$ compounds. Inset shows $C_p/T$ vs $T^2$ at different fields and solid red line is the fit to the relation $C_p(T) = \gamma T + \beta T^3 + \delta T^5$ (b) Electronic specific heat $C_e/T$ as a function of temperature under different magnetic field..

Fig. 10: Electronic structure along the line of high symmetry points (left panel) and density of states of $Pd_{1-x}Ni_xTe$:x=0, 0.25,05,0.75 and 1.0 respectively (a)-(e).

Fig. 1:

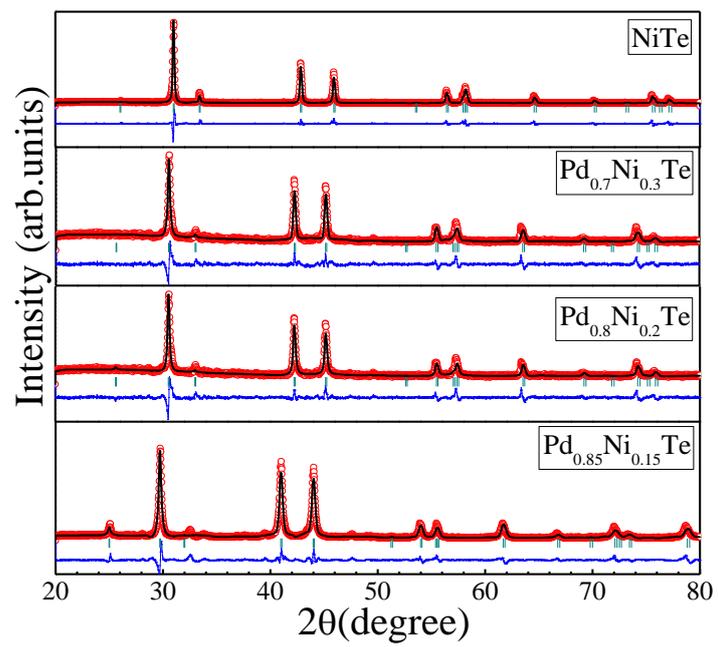

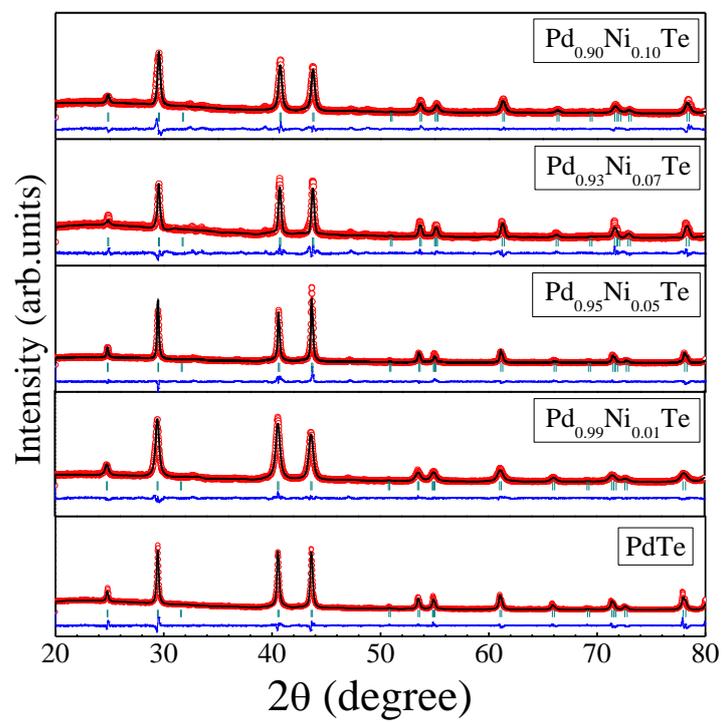

Fig. 2:

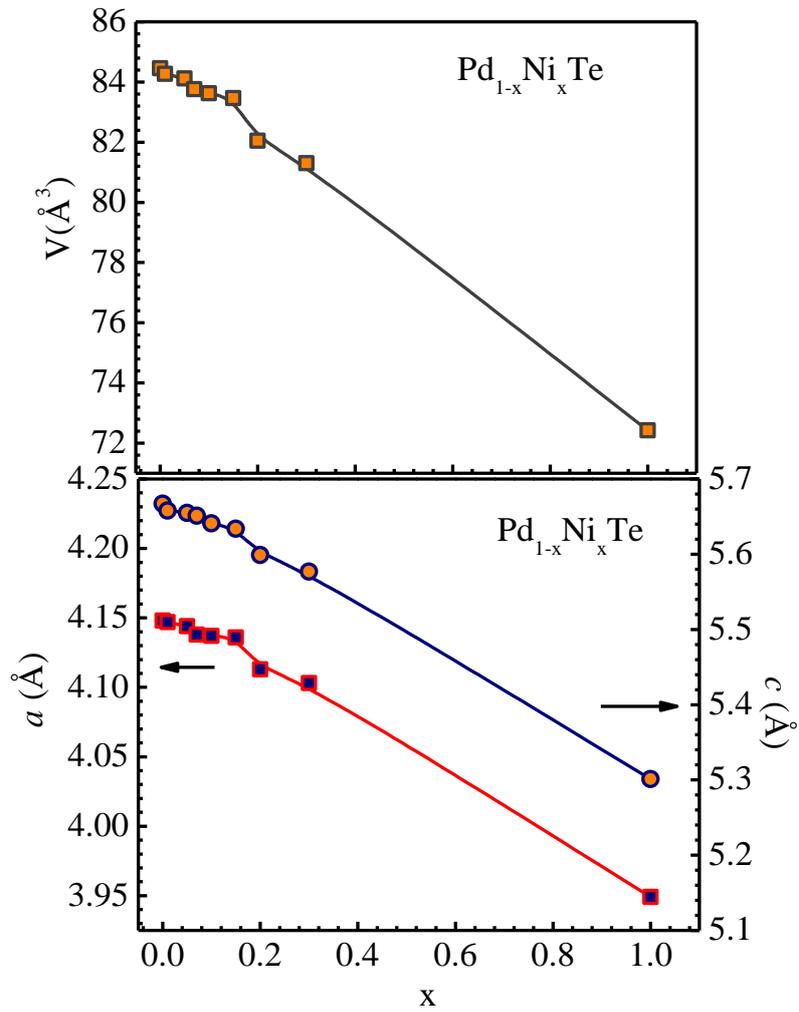

Fig. 3:

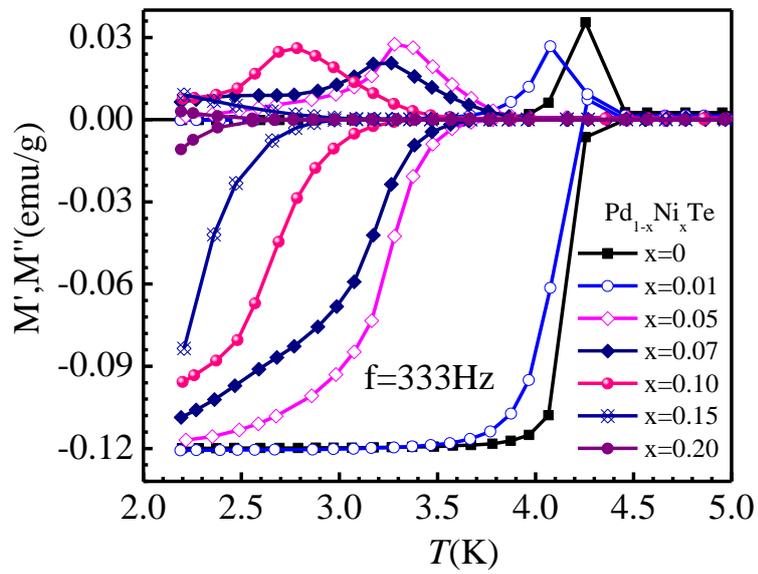

Fig. 4:

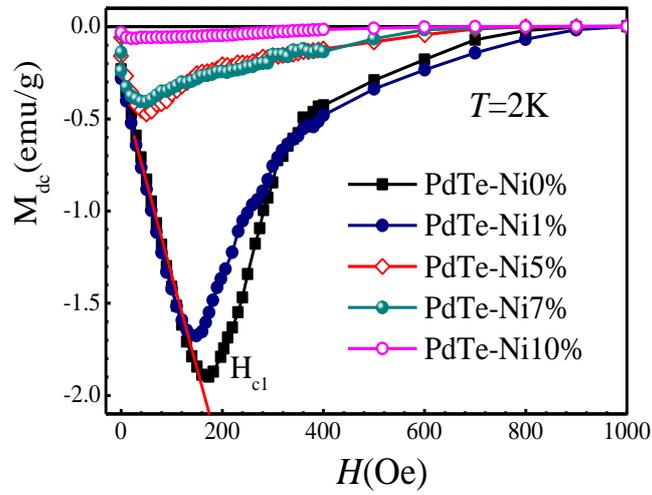

Fig.5:

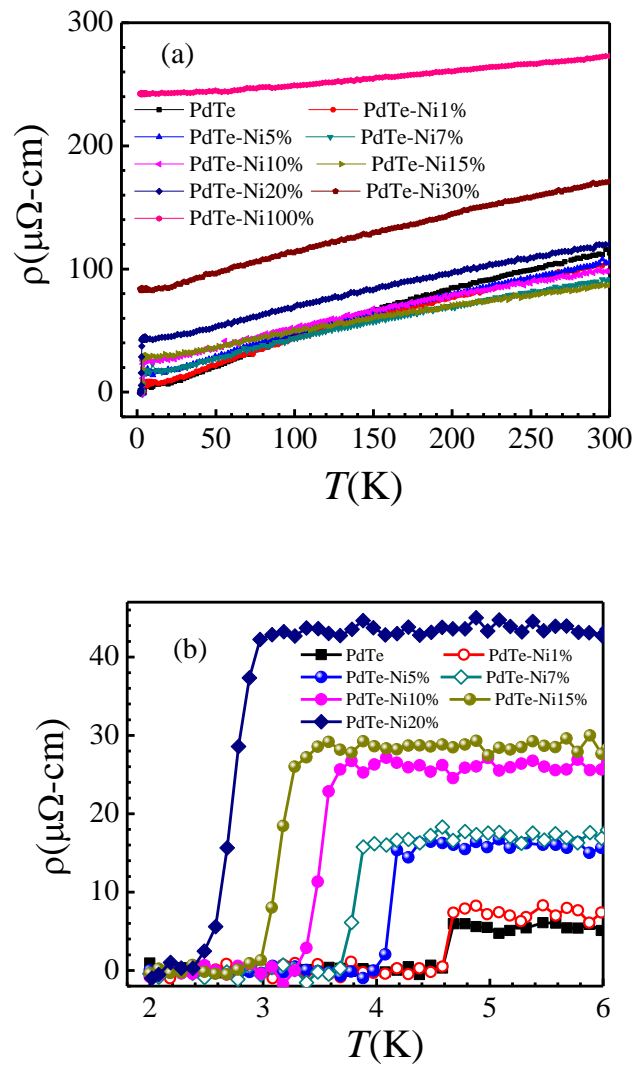

Fig. 6

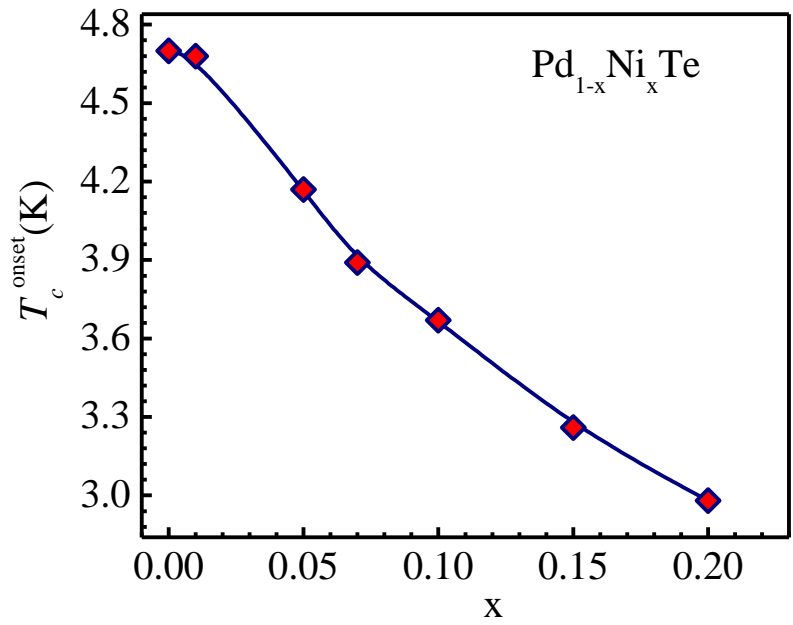

Fig 7:

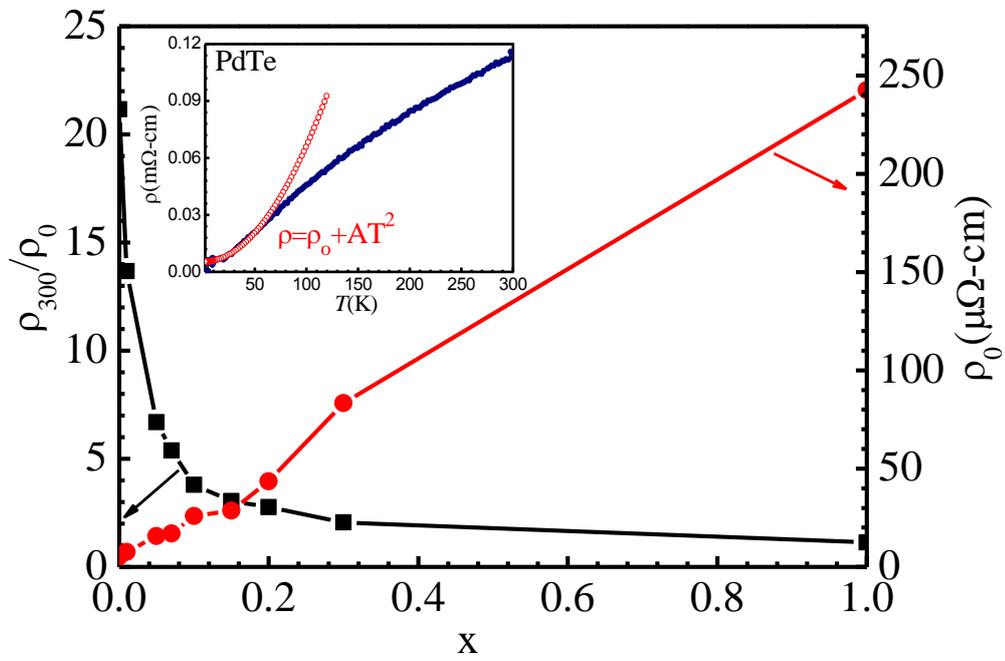

Fig.8:

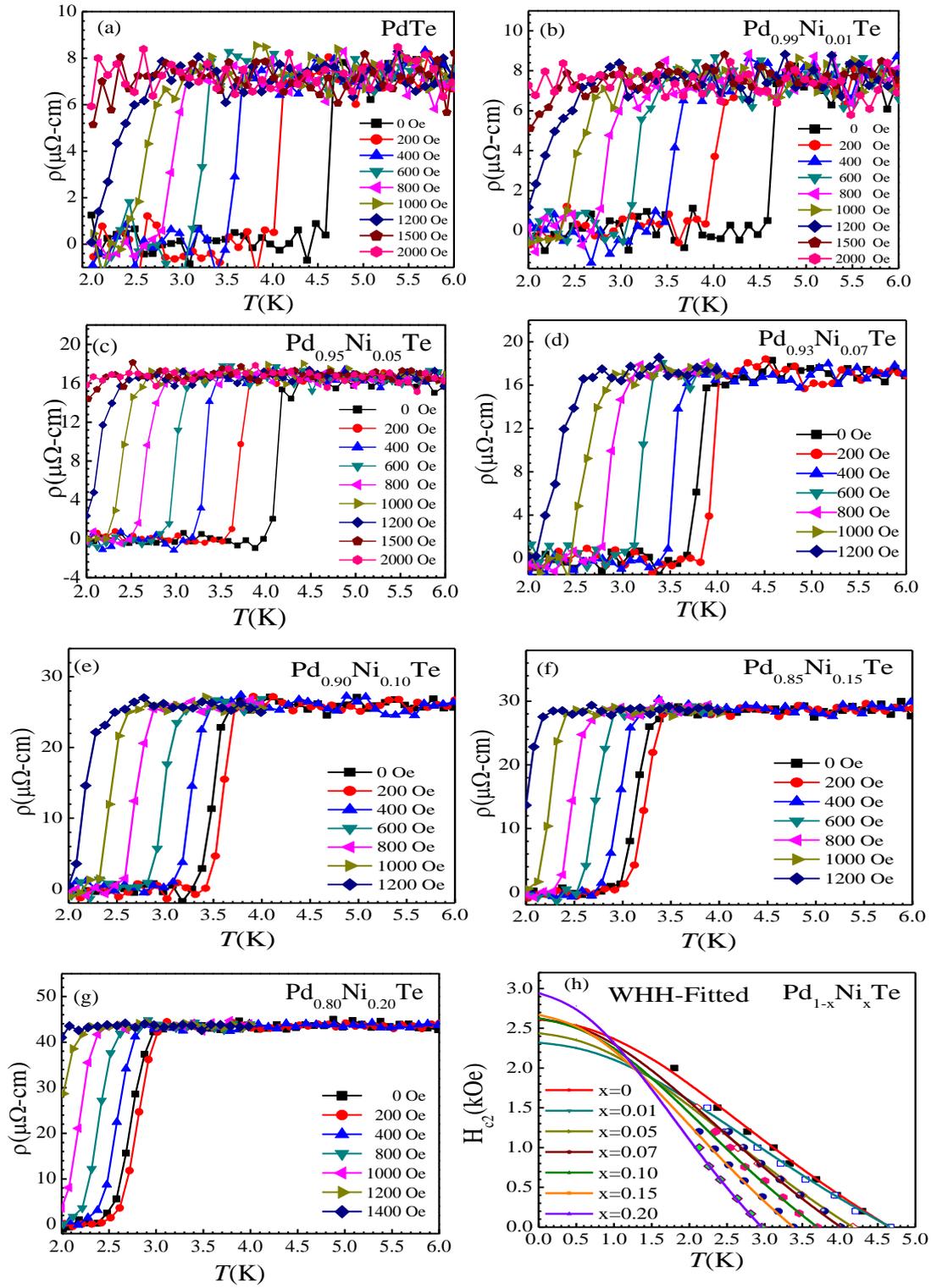

Fig 9:

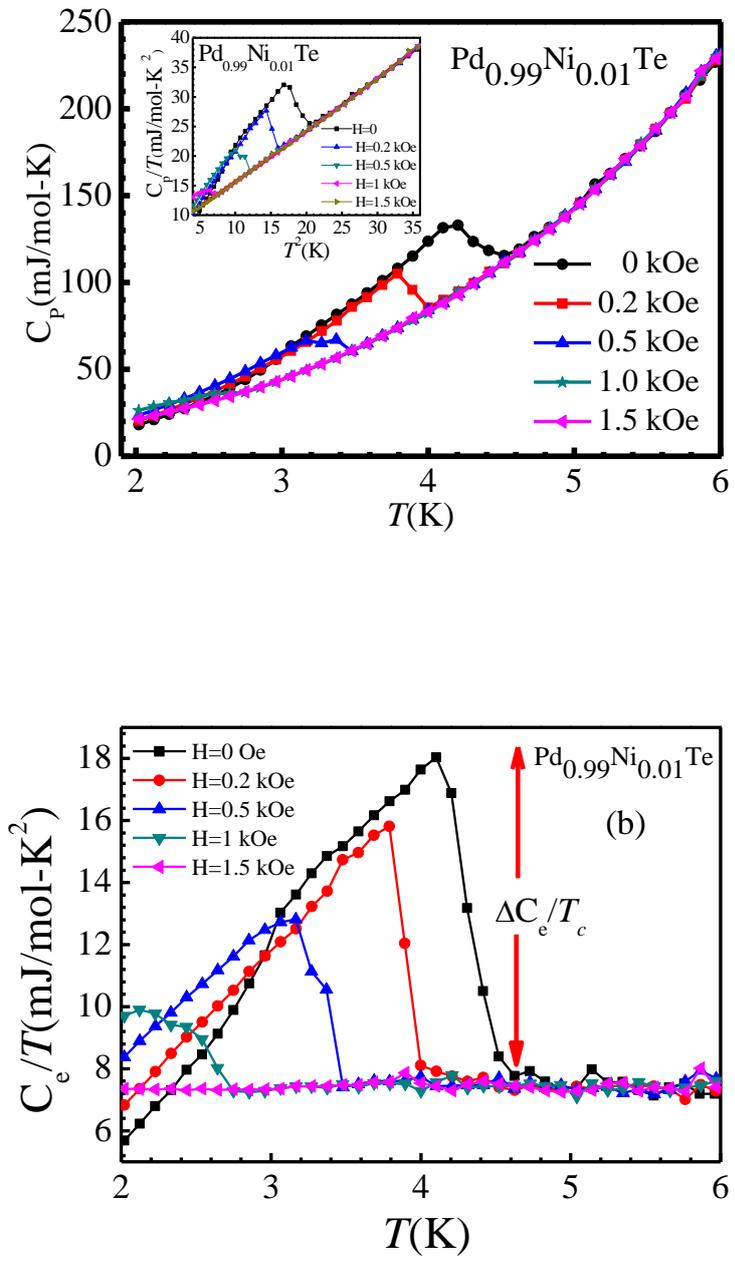

Fig 10:

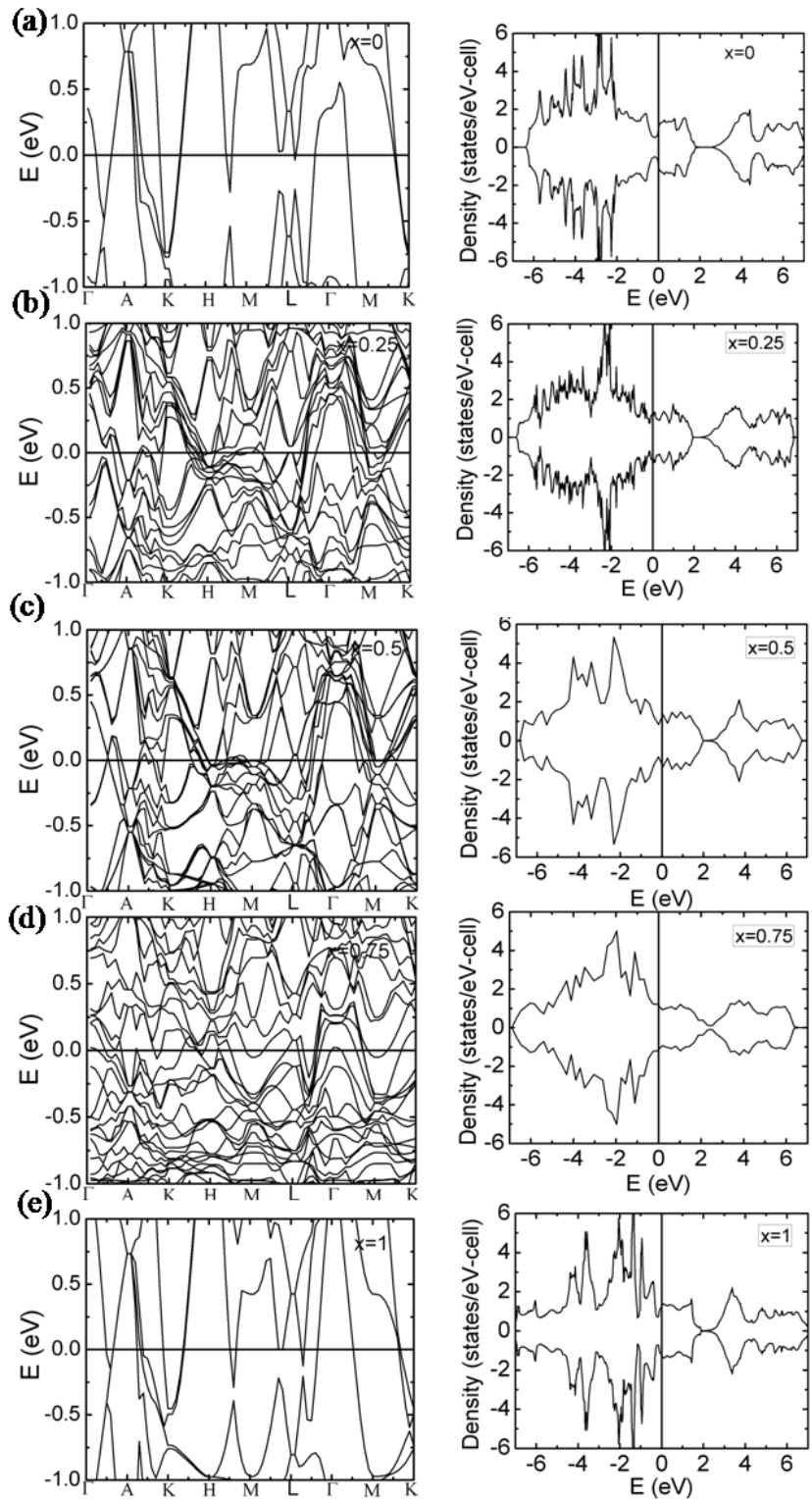